\documentclass[aps,prl,twocolumn,superscriptaddress,amsmath,amssymb]{revtex4}
\usepackage{graphicx}

\begin{document}

\date{\today}

\title{Multiplexing biochemical signals}

\author{Wiet de Ronde}\affiliation{FOM Institute for Atomic and Molecular Physics (AMOLF), Science Park 104, 1098 XG Amsterdam, The Netherlands}
\author{Filipe Tostevin}\affiliation{FOM Institute for Atomic and Molecular Physics (AMOLF), Science Park 104, 1098 XG Amsterdam, The Netherlands}
\author{Pieter Rein ten Wolde}\affiliation{FOM Institute for Atomic and Molecular Physics (AMOLF), Science Park 104, 1098 XG Amsterdam, The Netherlands}

\begin{abstract}
In this paper we show that living cells can multiplex biochemical signals,
{\em i.e.} transmit multiple signals through the same signaling pathway
simultaneously, and yet respond to them very specifically.
We demonstrate how two binary input signals can be encoded in the
concentration of a common signaling protein, which is then decoded such
that each of the two output signals provides reliable information about
one corresponding input. Under biologically relevant conditions the network
can reach the maximum amount of information that
can be transmitted, which is 2 bits.
\end{abstract}
\maketitle

\begin{figure}[b]
\begin{center}
\includegraphics[clip]{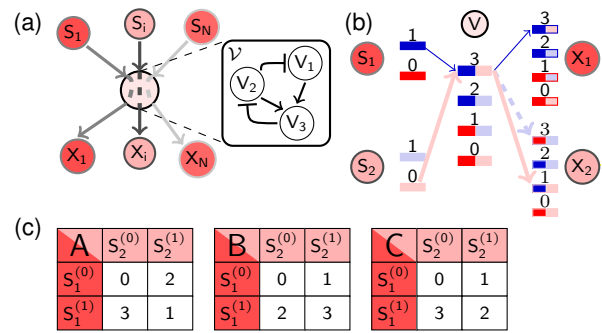}
\end{center}
\caption{(a) Biochemical multiplexing: $N$ different signals are
  encoded in the state of a common pathway ${\cal V}$, which is then
  decoded such that each output species ${\rm X}_i$ provides reliable
  information about the corresponding input ${\rm S}_i$. (b)
  Multiplexing is a mapping problem. The states of two inputs ${\rm
    S}_1$ and ${\rm S}_2$ are mapped onto the concentration of ${\rm
    V}$, which is then mapped onto states of the output species ${\rm
    X}_1$ and ${\rm X}_2$; we require that the two lowest (highest)
  levels of ${X}_i$ correspond to the lowest (highest) level of
  ${S}_i$; the dashed arrow denotes a mapping that violates this
  requirement; levels of ${\rm V}$ and ${\rm X}_i$ are colored
  according to input pattern ${\bf s}=(s_1,s_2)$. (c) The 3 unique mappings of ${\bf s}$ to $v$; in panel (b) mapping C is shown.}
\label{fig:cartoon}
\end{figure}

Cells continually have to respond to a myriad of signals.  One strategy for
transmitting distinct stimuli is to use distinct signal transduction networks .
It is, however, increasingly recognized that components are often shared
between pathways \cite{Schwartz:2004uq}.  Moreover, cells can transmit
different signals through one and the same pathway, and yet respond to them
specifically. In rat cells, for instance, neuronal growth factor and epidermal
growth factor stimuli are transmitted through the same MAPK pathway, yet give
rise to different cell fates, differentiation and proliferation respectively
\cite{Marshall:1995zr}. These observations suggest that cells are able to
transmit multiple messages through the same signal transduction network, just
as many telephone calls can be transmitted via a single wire. Indeed, the
intriguing question that arises is whether biochemical networks, like
electronic circuits, can {\em multiplex} signals: can multiple input signals be
combined (encoded) {\em simultaneously} in the dynamics of a common signalling
pathway, which are then decoded such that cells can 
respond specifically to each signal (see Fig.~\ref{fig:cartoon})?

The question of how cells can transduce multiple signals via pathways that
share components is a key question in biology, since sharing components may
lead to unwanted crosstalk between the different signals: from the perspective
of one signal, the presence of additional signals constitutes noise. In recent
years, several mechanisms for ensuring signaling specificity have been
proposed. One is spatial insulation, where the shared components are
incorporated into distinct macromolecular complexes on scaffold proteins
\cite{Schwartz:2004uq}. Other proposals are based on the temporal dynamics of
the system, such as cross-pathway inhibition
\cite{Bardwell:2007ly,McClean:2007lq,Hu:2009fk} and kinetic insulation
\cite{MarceloBehar10092007}. However, these studies only considered scenarios
in which the system is stimulated with one signal at the time. Rensing and
Ruoff studied what happens when two or three MAPK pathways that share
components are stimulated simultaneously \cite{Rensing:2009kx}, but found that
one pathway tends to dominate the response, suggesting that multiple messages
cannot be transmitted simultaneously. Here we demonstrate that cells can truly
multiplex signals: we show that they can transmit at least two signals
simultaneously through a common pathway, and yet respond specifically to each
of them.

We first have to understand how multiple signals can be encoded in the dynamics
of a signaling pathway. Cells employ a number of coding strategies for
transducing signals. One is to encode stimuli in the temporal dynamics, such as
the duration \cite{Marshall:1995zr} or frequency  \cite{Berridge:2000zr}, of an
intracellular signal.  In principle, these coding strategies could be used to
multiplex signals.  Here, we consider what is arguably the simplest and most
generic coding strategy cells could choose, namely one in which the signals are
encoded in the concentrations of the signaling proteins. We will call this
strategy AM multiplexing.

We will consider the biochemical network shown in Fig. \ref{fig:cartoon}A. It
consists of $N$ input species ${\rm S}_1,\dots,{\rm S}_{N}$ with copy numbers
$S_1, \dots,S_{N}$, a signal transduction pathway $\cal V$ consisting of $M$
species ${\rm V}_1,\dots,{\rm V}_{M}$, and $N$ output species ${\rm
X}_1,\dots,{\rm X}_{N}$. The copy number of each input species ${S}_i$ can be
in one of $K$ states, ${s}_i=0,\dots,K-1$, which are labelled in order of
increasing copy number, $S_i^{(0)}<S_i^{(1)}<\dots<S_i^{(K-1)}$. The input
pattern is denoted by the vector ${\bf {s}}=\left({s}_1,\dots,{s}_N\right)$.
Similarly, the copy number of each output species ${\rm X}_i$ can be in one of
$L$ states ${x}_i=0,\dots,L-1$ ordered by increasing copy number $X_i$, and the
output pattern is denoted by the vector ${\bf x}
=\left({x}_0,\dots,{x}_N\right)$. A necessary condition for multiplexing is
that the state space of $\cal V$ is large enough that it is possible to encode
the total number of input patterns, $K^N$, in $\cal V$.

We imagine that the $N$ input signals are independent, and that the signal
transduction network $\cal V$ replaces $N$ independent signaling pathways. We
therefore require that $X_i$ should provide reliable information about the
state ${s}_i$, but not necessarily about ${s}_{j\neq i}$; the $N$ different
input signals ${\bf s}$ simply have to be transduced to ${\bf x}$, not
necessarily integrated. In general, however, the state ${x}_i$ will be a
function of the states of all the input species: ${x}_i = f({\bf s})$. This
reflects the fact that inevitably there is cross-talk between the different
signals because they are transmitted via the same pathway. However, this
cross-talk is not detrimental as long as it does not compromise the cell's
ability to infer from ${x}_i$ what ${s}_i$ was.

Another key point is that while the precise mapping from ${\bf s}$ to ${\bf x}$
may not be critical for the amount of information transmitted {\em per se},
this is likely to be important for whether or not this information can be
exploited.  Let's imagine that the system contains three input species, say
three sugars, and that each of these can be in one of only two states,
${s}_i=0$ or $1$, corresponding to the absence or presence of the sugar; let's
further assume that ${\rm X}_i$ is an enzyme needed to consume sugar ${\rm
S}_i$.  With 8 input patterns $X_i$ can, in the absence of noise, take 8
values, identified as states ${x}_i=0,\dots,7$. Now, it seems natural to demand
that when the sugar ${\rm S}_i$ is absent (${s}_i=0$), the copy number of
enzyme ${\rm X}_i$ is low, while when ${\rm S}_i$ is present, the copy number
of ${\rm X}_i$ is high; this means that the four lowest levels of $X_i$
(${x}_i=0,1,2,3$) should correspond to ${s}_i=0$, while the four highest levels
of $X_i$ should correspond to ${s}_i=1$. We therefore require that the mapping
from ${\bf s}$ to ${\bf x}$ is such that the output states $\lbrace
{x}_i\rbrace$ corresponding to input ${s}_i=j$ are grouped into sets that are
{\em contiguous} and either increase or decrease {\em monotonically} with $j$,
for each signal $i$. This leads to a monotonic input-output relation between
$S_i$ and $X_i$ for each $i$. We call this requirement the multiplexing
requirement.

In the rest of the manuscript, we make these ideas concrete for a network in
steady state with two input species, ${\rm S}_1$ and ${\rm S}_2$, each of which
has either a low ($s_i=0$) or a high concentration ($s_i=1$). We take a
signaling pathway $\cal V$ consisting of only one species, ${\rm V}$.
Multiplexing requires that, in the absence of noise, the four input patterns
${\bf s}$ can be mapped onto four distinct states of $V$, ${v}=0,\dots,3$,
again labelled in order of increasing copy number. These four levels of $V$
lead to four states for each of the two output species ${\rm X}_1$ and ${\rm
X}_2$ (Fig.~\ref{fig:cartoon}B). As explained above, we require that we can
group these four states into two sets, called LOW and HIGH, such that the LOW
set, containing ${x}_i=0,1$, corresponds to ${s}_i=0$ and the HIGH set,
containing ${x}_i=2,3$, corresponds to ${s}_i=1$ (or {\em vice versa}, leading
to an inverse input-output relation). To elucidate which mechanisms make it
possible to multiplex ${\rm S}_1$ and ${\rm S}_2$, we note that there exists
different ways of mapping ${\bf s}$ to $v$, but, as we will explain shortly,
not all of these mappings can be decoded into ${\bf x}$ in a manner that
satisfies the multiplexing requirement. We therefore first address the question
which combinations of mapping from ${\bf s}$ to $v$ and decoding from $v$ to
${\bf x}$ fulfill the multiplexing requirement, and then we will discuss what
encoding mechanisms actually allow for the required mapping from ${\bf s}$ to
$v$.

Due to the symmetry in the problem, there are 3 unique ways of mapping the four
input patterns ${\bf s}$ to $v$ (Fig. \ref{fig:cartoon}C). To determine whether
there exists a scheme for decoding the signals from $v$ to ${\bf x}$ that
satisfies the multiplexing requirement, we examine for each mapping all
possible network topologies between ${\rm V}$, ${\rm X}_1$ and ${\rm X}_2$,
except those that involve autoregulation or mutual repression/activation since
these may lead to bistability. In particular, we allow not only for activation
and repression of ${\rm X}_1$ and ${\rm X}_2$ by ${\rm V}$, but also for
activation and repression of ${\rm X}_2$ by ${\rm X}_1$, leading to feedforward
loops, a common motif in signal transduction pathways and gene networks
\cite{Shen-Orr:2002uq}. In the deterministic mean-field limit the steady-state
values of $X_1$ and $X_2$ are thus given by
\begin{eqnarray}
	X_1&=&k_1 f(V;K_\alpha,n_\alpha) /\mu, \label{eq:X1}\\
	X_2 &=&k_2 f(V;K_\beta,n_\beta)  \times f
  	(X_1;K_\gamma,n_\gamma) /\mu,\label{eq:ffl} \label{eq:X2}
\end{eqnarray}
where $k$ is the maximum activation/production rate, $\mu$ is the
degradation/deactivation rate, and each regulation function is either an
activating or repressing Hill function, $f(V;K,n)={V^n}/({V^n+K^n})$ or
$f(V;K,n)={K^n}/({V^n+K^n})$. The multiplication in Eq.~\ref{eq:ffl} indicates
that we assume that at $X_2$, $X_1$ and $V$ are integrated according to AND
logic \cite{Shen-Orr:2002uq}.
We performed extensive sampling of the space of parameters $k_1,
k_2, K_\alpha,n_\alpha,K_\beta,n_\beta,K_\gamma,n_\gamma$ for each of the
mappings in Fig.~\ref{fig:cartoon}C.

\begin{figure}[tb]
\begin{center}
\includegraphics[clip]{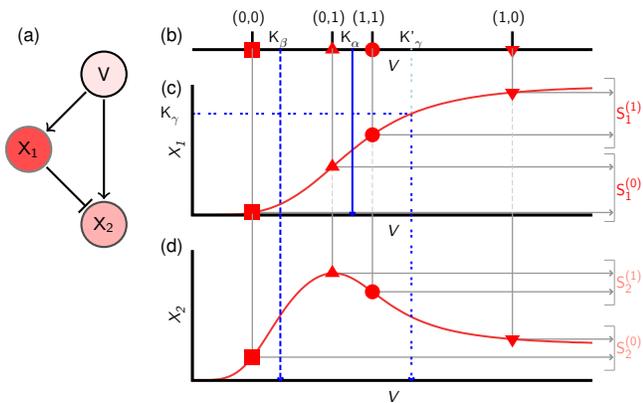}
\end{center}
\caption{Decoding $V$ using an incoherent feedforward loop. (a) Network
  architecture.  (b) The values of $V$ corresponding to the four input
  patterns ${\bf s}$ for mapping C (see Fig. \ref{fig:cartoon}C), with
  thresholds $K_\alpha, K_\beta, K_\gamma$ (see
  Eqs. \ref{eq:X1}-\ref{eq:X2}). (c) $X_1(V)$; (d) $X_2(V)$. The
  non-monotonicity of $X_2(V)$ swaps the states corresponding to
  $(1,1)$ and $(0,1)$ in the mapping from $v$ to $x_2$.}
\label{fig:decoding}
\end{figure}

Only for mapping C do we find decoding schemes that satisfy the multiplexing
requirement
\footnote{Mapping B does allow valid multiplexing solutions, but
	only with unrealistic parameter values.}.
Interestingly, all valid decoding networks are incoherent feedforward loops
\cite{Shen-Orr:2002uq}. Figure~\ref{fig:decoding} illustrates the principle
for one such motif. Panel B shows for each of the four input patterns ${\bf
s}$ the copy number $V$ together with the threshold copy numbers, $K_\alpha$,
$K_\beta$ and $K_\gamma$, while panels C and D show $X_1$ and $X_2$
respectively as a function of $V$.  $X_1(V)$ is a simple activation curve with
activation threshold $K_\alpha$. In contrast, $X_2(V)$ starts low and rises
around $K_\beta$, but then decreases again due to repression by ${\rm X}_1$.
This non-monotonicity, which is a result of the incoherent character of the
feedforward loop, is critical, since this makes it possible to swap the order
of the states corresponding to ${\bf s}=(1,1)$ and $(1,0)$ in the mapping from
$v$ to $x_2$. The key parameters are the activation/repression thresholds
$K$, since they determine where in the state space of $V$ the outputs switch
between high and low levels. The precise values of $k$ and $n$ are of less
importance, although $n$ should not become so large that $X_1(V)$ becomes
Boolean: it is critical that ${\rm X}_1$, which needs to be activated by ${\rm
V}$ around $K_\alpha$ to transmit $S_1$, is not fully activated at $K_\alpha$:
to multiplex $S_2$, $X_1$ should reach the threshold $K_\gamma$ for repressing
${\rm X}_2$ only when $V$ has become significantly larger than $K_\alpha$.
Indeed, if $X_1$ can only take two states, then only three states of $V$ could
be decoded, and not the required four. AM multiplexing thus relies on the fact
that signals can be encoded over a range of concentrations. 

We can now also
understand why mappings A and B are difficult to decode: they would require an
input-output relation between $X_2$ and $V$ that rises more than once. This is
difficult to achieve in a feedforward loop without mutual repression or
activation.

\begin{figure}[t]
\begin{center}
\includegraphics[clip]{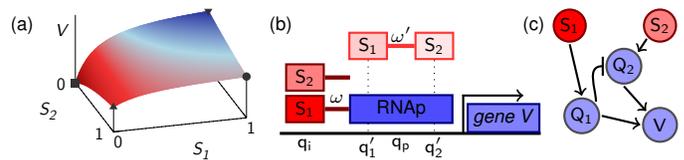}
\end{center}
\caption{(a) An input-output relation $V(S_1,S_2)$ consistent with
  mapping C (Fig. \ref{fig:cartoon}C); symbols $\blacksquare,
  \blacktriangle, \bullet, \blacktriangledown$ correspond to states in
  Fig.~\ref{fig:decoding}. (b) Architecture of the promoter. (c) A
  feedforward loop that can generate mapping C.}
\label{fig:promoter}
\end{figure}

The above analysis shows that it is possible to decode multiple signals
simultaneously, provided that the input ${\bf s}$ can be encoded in $v$
according to mapping C. The next question is how these mappings, which
correspond to particular input-output relations $V(S_1,S_2)$, can be generated.
Experiments \cite{Kaplan:2008vn} and modelling \cite{Buchler03,Hermsen:2006fj}
have shown that transcriptional regulation can be very sophisticated, allowing
for complex logical operations \cite{Hermsen:2006fj}. We indeed find that a
simple scheme for transcriptional regulation based on the mechanism of
`regulated recruitment' \cite{Buchler03} can generate the required
input-output relation $V(S_1,S_2)$, where ${\rm S}_1$ and ${\rm S}_2$ are now
transcription factors that regulate the expression of the protein ${\rm V}$. In
this scheme, ${\rm S}_1$ and ${\rm S}_2$ independently activate gene expression
by binding next to the core promoter, thus recruiting the RNA polymerase
(RNAp), while ${\rm S}_1$ and ${\rm S}_2$ together repress gene expression by
cooperative binding to the core promoter, thereby blocking the binding of RNAp
(see Fig.~\ref{fig:promoter}). This yields
\begin{equation}
\label{eq:V}
V(S_1,S_2) = \frac{(\beta/\mu) \,q_p(1+\omega q_1 + \omega q_2)}{1 + q_1^\prime +
  q_2^\prime + \omega^\prime q_1^\prime q_2^\prime + q_p(1+\omega q_1 +
\omega q_2)},
\end{equation}
where $\beta$ is the maximum expression rate and $\mu$ is the degradation rate
of ${\rm V}$, $q_p = c_p/K_p$ is the concentration of RNAp $c_p$ scaled with
its dissociation constant $K_p$, $q_1=S_1/K_1$, $q_2=S_2/K_2$,
$q_1^\prime=S_1/K_1^\prime$, and $q_2^\prime=S_2/K_2^\prime$, where $K_i$ and
$K_i^\prime$ are the dissociation constants  for the binding of ${\rm S}_i$ to
the promoter sites where the RNAp is recruited or blocked, respectively;
$\omega$ and $\omega^\prime$ are factors reflecting cooperative interactions between the
respective molecules \cite{Buchler03,Hermsen:2006fj}. We thus conclude that
gene regulation networks have the capacity to multiplex signals.

While it is clear that signaling pathways often share common
components \cite{Schwartz:2004uq,Marshall:1995zr}, the logic of signal
integration in these pathways has been characterized in much less detail than
for gene regulatory networks. It is conceivable that the desired input-output
function $V(S_1,S_2)$ could be implemented at the level of a single protein
${\rm V}$, using competitive and/or cooperative binding between the three
molecules ${\rm S}_1$, ${\rm S}_2$, ${\rm V}$. Alternatively, the required
encoding could also be implemented at a higher level of network interactions.
For instance, a network in which ${\rm S}_1$ and ${\rm S}_2$ regulate $V$ via
two additional components, ${\rm Q}_1$ and ${\rm Q}_2$, in an incoherent
feedforward loop (Fig.~\ref{fig:promoter}C), could achieve the required
encoding $V(S_1,S_2)$.  In essence, the feedforward loop between ${\rm Q}_1$,
${\rm Q}_2$ and ${\rm V}$ can be used to control the ordering of $V$ in the
encoding process, just as the feedforward loop between ${\rm V}$, ${\rm X}_1$
and ${\rm X}_2$ can be used to regulate the ordering of $X_2$ in the decoding
step. Since feedforward loops are common motifs in signal transduction
pathways \cite{Shen-Orr:2002uq}, we argue that multiplexing can also be
implemented in these networks.

\begin{figure}[t]
\begin{center}
\includegraphics[clip]{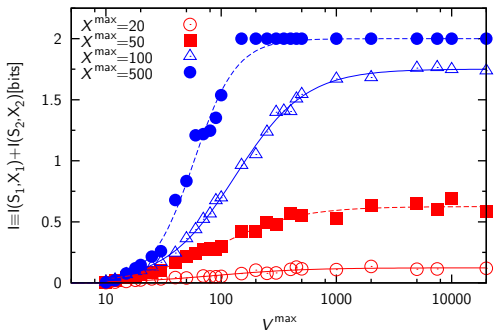}
\end{center}
\caption{The transmitted information
  $I$ as a function of $V^{\rm max}$ for four different values of $X^{\rm
    max}$.  $I$ can reach the maximum value of 2 bits provided $V^{\rm
    max}$ and $X^{\rm max}$ are large enough.}
\label{fig:I_V}
\end{figure}

The analysis above shows that in principle biochemical networks can multiplex
signals in the mean-field, deterministic limit. However, there remains the
question of whether signals can be multiplexed reliably in the presence of
inevitable biochemical noise. To address this, we estimate a lower bound on the
information about two binary signals $S_1$ and $S_2$ that are transmitted
through the network studied above (Eqs.~\ref{eq:X1}-\ref{eq:V}). We define the
total information $I\equiv I(S_1,X_1)+I(S_2,X_2)$ as the sum of the mutual
information for each of the individual signals,
$I(S_i,X_i)=\sum_{x_i}\sum_{s_i}p(x_i,s_i)\log [p(x_i,s_i)/p(x_i)p(s_i)]$
\cite{Shannon48}, where $p(s_i)$ and $p(x_i)$ are respectively the
probabilities of $S_i$ being in state $s_i$ and $X_i$ being in state $x_i$, and
$p(x_i,s_i)$ is the joint probability of input $s_i$ and output $x_i$. Note
that in the presence of noise $X_i$ is not limited to 4 states but can in
principle take any value. This definition of $I$ makes it straightforward to
directly compare the performance of this network with that of two independent
pathways. If each of the two input states for each $S_i$ is equally likely
then the maximum value of $I(S_i,X_i)$ is 1 bit for each transmitted signal
$i$; the maximum value of $I$ is thus 2 bits.

To maximize the lower bound on $I$ we optimize the network parameters using a
simulated-annealing algorithm; we have verified that the final results are
robust by varying the initial conditions, and by also using an evolutionary
algorithm. We fix the degradation rate of all proteins to be $\mu=1 {\rm
hr}^{-1}$ and vary $n_\alpha$, $n_\beta$ and $n_\gamma$ between 1 and 4. Values
of $k_1$, $k_2$ are set such that the maximum mean value of each $X_i$ is
$X^{\rm max}$; similarly, $\beta$ is set such that the maximum
mean value of $V$ is $V^{\rm max}$. $X^{\rm max}$ and $V^{\max}$ are varied
systematically (see Fig.  \ref{fig:I_V}). The threshold parameters
$K_{\alpha},K_{\beta}$ and $K_{\gamma}$ are varied over the range $[0,V^{\rm
max}]$ or $[0,X^{\rm max}]$ as appropriate. We vary $q_p$, $q_i$ from $10^{-2}$
to $10^2$ and $\omega$, $\omega'$ between $1$ and $10$. For each parameter set
we compute $p(x_i,s_i)$ using the linear-noise approximation \cite{kampenbook}.
Its accuracy was verified by performing Gillespie
simulations of the optimized networks \cite{Gillespie77}.

Figure~\ref{fig:I_V} shows that below a threshold copy number $V^{\rm
max}_c\approx50$ the total information is low regardless of $X^{\rm max}$
because four distinct states of $V$ cannot be generated. Above $V^{\rm max}_c$,
for large $X^{\rm max}$ the information $I$ reaches 2 bits, the maximum
information about the two signals $S_1$ and $S_2$ that could be transmitted via
two independent channels. For lower values of $X^{\rm max}$, $I$ saturates at
a value lower than 2 bits, limited by the intrinsic noise in the production and
decay of ${\rm X}_i$. Importantly, $I$ reaches 2 bits for $V^{\rm max}\approx
X^{\rm max}\approx 500$, which is well within the range of typical protein copy
numbers inside living cells. This shows that biochemical networks can multiplex
two signals reliably in the presence of biochemical noise under biologically
relevant conditions.

In summary, our results suggest that cells can transmit at least two binary
signals through one and the same pathway, and yet respond specifically and
reliably to each of them. The proposed mechanism for biochemical multiplexing
is based on swapping the order of states during the encoding and decoding steps
using incoherent feedforward loops. It is clear that the principle is generic,
and could be implemented in signal transduction pathways and gene networks --
indeed incoherent feedforward loops are commonly found in these networks
\cite{Shen-Orr:2002uq}. Our predictions could be tested experimentally by
simultaneously stimulating two MAPK pathways that share components
\cite{Schwartz:2004uq}, although perhaps a more controlled experiment would be
one using synthetic gene networks. In future work, we will address how more
than two input signals can be transduced simultaneously, and how cells can
multiplex signals by encoding them into the temporal dynamics of the signaling
pathway.

We thank Tom Shimizu for a critical reading of the manuscript. This work is
supported by FOM/NWO.

\end{document}